\documentstyle[12pt]{article}
\begin{document}
{Galactic dynamo seeds from non-superconducting spin-polarised
strings} \vspace{1cm}\begin{center} \noindent L.C.Garcia de
Andrade\footnote{Departamento de Fisica Teorica,Instituto de
F\'{\i}sica , UERJ, Rua S\~{a}o francisco Xavier 524, Rio de
Janeiro,CEP:20550-013, Brasil.e-mail:garcia@dft.if.uerj.br.}
\end{center}
\vspace{1cm}
\begin{center}
{\Large Abstract}
\end{center}
\vspace{0.5cm} Earlier Enqvist and Olesen have shown that formation
of ferromagnetic planar walls in vacuum at GUT scales in comoving
plasmas may generate a large scale magnetic field of
$B_{now}\simeq{10^{-14}G}$. In this paper we show that starting from
classical Einstein-Cartan-Maxwell strong gravity, a spin-polarised
ferromagnetic cylinder gives rise to a cosmological magnetic field
of the order $B_{now}\simeq{10^{-22}G}$. Vorticity of cylinder is
used to obtain galactic magnetic fields. Magnetic fields up to
$B\sim{10^{9}G}$ can be obtained from the spin density of the
cylinder. If matching conditions are used cosmological magnetic
fields of the order of $B\sim{10^{-16}R\frac{Gauss}{cm}}$ where $R$
is the radius of the cosmic strings. For a cosmic string with the
radius of an hydrogen atom the cosmic magnetic field is
$B\sim{10^{-32}Gauss}$ which is enough to seed galactic dynamos.
Current of these cosmic strings are computed and its shown that
these strings are non-conducting since the electric current is much
weaker than the one produced by string dynamos. Taking into account
vorticity of the cosmic string one obtains a B-field which coincides
with the IGM field of $10^{-5}Gauss$.
\newpage
\section{Introduction}
Earlier the author has published a paper in CQG \cite{1} where
spin-polarised cylinders with magnetic fields were used as a tool to
test Einstein-Cartan theory of gravitation \cite{2}. In the present
paper we compute from the negative pressure exact solutions of ECM
equations two expressions one for the magnetic field in terms of
torsion which we show to yield a B-field of $10^{-16}G$ compatible
with the values estimated by Barrow et al \cite{3} using only
general relativity plus usual Maxwell equations, and the other of
the magnetic field in terms of the vorticity of the spin-polarised
cylinder which also produces the ${\mu}G$ galactic magnetic fields
as seen for example in the Milky way. Both results shows that ECM
gravity is compatible with the observed astronomical data of dynamo
mechanism not only for galactic dynamos but also for the
cosmological magnetic fields as seeds for these galactic dynamos.
This shows also that the recently present author efforts \cite{4} to
show that torsion of spacetime could be compatible with dynamo
mechanism and primordial magnetic fields. When matching conditions
are used on spin-polarised non-rotating cylinders one obtains a
relation between the large scale magnetic fields and the radius of
the cosmic strings. For strings of radius of scale of hydrogen atom
B-fields as low as $10^{-24}Gauss$ are found. In section 2 we
discuss the magnetogenesis of the cosmic string in terms of the
matching conditions in Riemann-Cartan spacetime. In section 3
instead of this vorticity free case we consider the Harrison-Rees
vorticity and compute the magnetic field from the spin-polarised
cosmic string which yields the IGM of $10^{-5}Gauss$.
\section{Matching conditions for cosmic strings in torsioned spacetime and magnetogenesis}
Recently we have shown \cite{5} that Soleng \cite{6} cylinder
geometry given by
\begin{equation}
ds^{2}=-(e^{\alpha}dt+Md{\phi})^{2}+r^{2}e^{-2{\alpha}}d{\phi}^{2}+e^{2{\beta}-2{\alpha}}(dr^{2}+dz^{2}).
\label{1}
\end{equation}
in the particular case of ${\alpha}={\beta}=0$ can be shown to be a
solution of Einstein-Cartan-Maxwell (ECM) field equations which
represents a magnetized spin polarised cylinder in torsioned
spacetime where the RC rotation vanishes when one applies the
matching condition in this non-Riemannian space. $M$ is a function
of the radial coordinate $r$. Exterior solution is the same as used
by Soleng in the case of thick spinning cosmic strings \cite{6} and
represents an exterior solution of Einstein's vacuum field equation
\begin{equation}
R_{ik}=0 \label{2}
\end{equation}
$(i,j=0,1,2,3)$ as
\begin{equation}
ds^{2}=-dt^{2}-2adtd{\phi}+dr^{2}+(B^{2}(r+r_{0})^{2}-a^{2})d{\phi}^{2}+dz^{2}
\label{3}
\end{equation}
Here $a$, B and $r_{0}$ are constants. Before proceed in this
analysis let us consider the above metric (\ref{1}) in terms of the
differential one-form basis
\begin{equation}
{\theta}^{0}=e^{\alpha}dt+Md{\phi} , \label{4}
\end{equation}
\begin{equation}
{\theta}^{1}=e^{{\beta}-{\alpha}}dr , \label{5}
\end{equation}
\begin{equation}
{\theta}^{2}=re^{-{\alpha}}d{\phi} , \label{6}
\end{equation}
\begin{equation}
{\theta}^{3}=e^{{\beta}-{\alpha}}dz . \label{7}
\end{equation}
Polarisation along the axis of symmetry is considered and the Cartan
torsion is given in terms of differential forms by
\begin{equation}
T^{i}=2k{\sigma}{\delta}^{i}_{0}{\theta}^{1}{\wedge}{\theta}^{2}
\label{8}
\end{equation}
where ${\sigma}$ is a constant spin density. For computational
convenience we addopt Soleng's definition \cite{6} for the RC
rotation ${\Omega}$
\begin{equation}
{\Omega}:= -\frac{1}{2}{\sigma}+\frac{M'}{2r} \label{9}
\end{equation}
where ${\Omega}$ is the cylinder RC vorticity. Cartan's first
structure equation is
\begin{equation}
T^{i}=d{\theta}^{i}+{{\omega}^{i}}_{k}{\wedge}{\theta}^{k}
\label{10}
\end{equation}
and determines the connection forms ${{\omega}^{i}}_{j}$.The
connection one-forms are given by
\begin{equation}
{\omega}^{0}_{1}= -{\Omega}{\omega}^{2} \label{11}
\end{equation}
\begin{equation}
{\omega}^{0}_{2}={\Omega}{\omega}^{1} \label{12}
\end{equation}
\begin{equation}
{\omega}^{0}_{3}=0 \label{13}
\end{equation}
\begin{equation}
{\omega}^{1}_{2}=-{\Omega}{\omega}^{0}-(\frac{1}{r}){\omega}^{2}
\label{14}
\end{equation}
while others vanish. From the Cartan's second structure equation
\begin{equation}
{R^{i}}_{j}=d{{\omega}^{i}}_{j}+{{\omega}^{i}}_{k}{\wedge}{{\omega}^{k}}_{j}
\label{15}
\end{equation}
where the curvature RC forms
${R^{i}}_{j}={R^{i}}_{jkl}{\theta}^{k}{\wedge}{\theta}^{l}$ where
${R^{i}}_{jkl}$ is the RC curvature tensor. This is accomplished by
computing the RC curvature components from the Cartan structure
equations as
\begin{equation}
R_{0101}={\Omega}^{2} , \label{16}
\end{equation}
\begin{equation}
R_{0112}={\Omega}' , \label{17}
\end{equation}
\begin{equation}
R_{0202}={\Omega}^{2} , \label{18}
\end{equation}
\begin{equation}
R_{1201}={\Omega}' , \label{19}
\end{equation}
\begin{equation}
R_{1212}=3{\Omega}^{2}-2{\Omega}{\sigma} , \label{20}
\end{equation}
others zero. The dash here represents the derivative $w.r.t$ to the
radial coordinate $r$. From the curvature expressions above it is
possible to built the ECM field equations as
\begin{equation}
{G^{i}}_{k}= k{T^{i}}_{k} \label{21}
\end{equation}
where ${G^{i}}_{k}$ is the Einstein-Cartan tensor and ${T^{i}}_{k}$
is the total energy-momentum tensor composed of the fluid tensor
${T^{i}}_{k}=({\rho},p_{r},0,p_{z})$ and the electromagnetic field
tensor
\begin{equation}
{t^{i}}_{k}= ({F^{i}}_{l}
{F^{l}}_{k}-\frac{1}{2}{\delta}^{i}_{k}(E^{2}-B^{2})) \label{22}
\end{equation}
where $F_{0{\gamma}}$ correspond to the electric field $\vec{E}$
while $F_{{\alpha}{\beta}}$ components of the Maxwell tensor field
$F_{ij}$ correspond to the magnetic field $\vec{B}$. Here we
consider that the electric field vanishes along the cylinder, and
${{\alpha}=1,2,3}$. Thus the natural notation $E^{2}= (\vec{E})^{2}$
and the same is valid for the magnetic field. Thus explicitly the
ECM equations read
\begin{equation}
-3{\Omega}^{2}-{\sigma}{\Omega}=-k({\rho}+\frac{B_{z}}{2})
\label{23}
\end{equation}
\begin{equation}
{\Omega}^{2}=k(p_{r}-\frac{B_{z}}{2}) \label{24}
\end{equation}
\begin{equation}
{\Omega}'=0 \label{25}
\end{equation}
\begin{equation}
{\Omega}^{2}+{\sigma}{\Omega}=-k(p_{z}+\frac{B_{z}}{2}) \label{26}
\end{equation}
Note that equation (\ref{25}) is the simplest to solve and yields
${\Omega}={\Omega}_{0}= constant$. This actually from the Riemann
curvature expressions above shows that the Riemann tensor vanishes
outside the cylinder shell string which shows that from the metrical
point of view the spacetime is flat or Minkowskian. Therefore so far
just from the ECM field equations we cannot say that the cylinder is
static. Before proceed therefore it is useful to show that this
results from the Arkuszewski-Kopczynski-Ponomariev (AKP) \cite{7}
junction conditions for Einstein-Cartan gravity which match an
interior solution of ECM field equations to the exterior vacuum
solution given by the geometry given by expression (\ref{3}). The
AKP conditions are
\begin{equation}
{g}_{ij,r}|_{+}=g_{ij,r}|_{-} - 2K_{r(ij)} \label{27}
\end{equation}
for $(i,j)$ distinct from the r coordinate, where the contortion
tensor is
\begin{equation}
K_{ijk}= \frac{1}{2}( T_{jik}+T_{jki}-T_{ijk} ) \label{28}
\end{equation}
where $T_{jik}$ is the Cartan torsion. The plus and minus signs here
correspond respectively to the exterior and interior spacetimes
respectively. The others AKP conditions state that the fluid
elements do not move across the junction surface, the stress normal
to the junction surface vanishes and that
\begin{equation}
{g}_{ij}|_{+}=g_{ij}|_{-} \label{29}
\end{equation}
which is the general relativistic Lichnerowicz condition. From the
cylinder geometry one obtains
\begin{equation}
g_{t{\phi}}|_{+}= g_{t{\phi}}|_{-} \label{30}
\end{equation}
\begin{equation}
{g}_{{\phi}{\phi}}|_{+}=g_{{\phi}{\phi}}|_{-} \label{31}
\end{equation}
\begin{equation}
g_{t{\phi},r}|_{+}= g_{t{\phi},r}|_{-} - T_{tr{\phi}} \label{32}
\end{equation}
\begin{equation}
g_{{\phi}{\phi},r}|_{+}= g_{{\phi}{\phi},r}|_{-} - 2
T_{{\phi}r{\phi}} \label{33}
\end{equation}
which for the exrerior and interior of the cylinder matching at
$r=R$ one obtains
\begin{equation}
a=M(R) \label{34}
\end{equation}
\begin{equation}
B^{2}(R+r_{0})^{2}= R^{2}- M^{2} \label{35}
\end{equation}
\begin{equation}
{\Omega}_{0}= -\frac{1}{2}{\sigma}_{0}+\frac{M'}{2R} \label{36}
\end{equation}
\begin{equation}
0= {\sigma}_{0}R-M' \label{37}
\end{equation}
\begin{equation}
B^{2}(R+r_{0})= R -MM'+MR{\sigma}_{0} \label{38}
\end{equation}
Substitution of $M'$ above into expression (\ref{36}) yields the
desired result that the Riemann-Cartan rotation ${\Omega}$ vanishes.
The remaining junction conditions \cite{7} yield
\begin{equation}
B^{2}=\frac{R}{(R+r_{0})} \label{39}
\end{equation}
\begin{equation}
{\sigma}_{0}= \frac{4}{R}[1-\frac{R+r_{0}}{R}] \label{40}
\end{equation}
Substitution of these results into the exterior metric yields the
following exterior spacetime for the spin polarised cylinder
\begin{equation}
ds^{2}=-dt^{2}-2{\sigma}_{0}R^{2}dtd{\phi}+dr^{2}+R(\frac{(r+r_{0})^{2}}{R+r_{0}}-{{\sigma}_{0}}^{2}R^{3})d{\phi}^{2}+dz^{2}
\label{41}
\end{equation}
Now going back to the ECM equations we obtain the following
constraints
\begin{equation}
{\rho}=\frac{B_{z}}{2} \label{42}
\end{equation}
which states that the energy density is purely of magnetic origin.
Besides to keep the stability of the spin polarised cylinder and its
static nature one obtains that the radial pressure $p_{r}>0$ while
the axial pressure is negative or $p_{z}<0$. Indeed from the field
equations we obtain $p_{z}= - \frac{{B_{z}}^{2}}{2}$ while $p_{r}=
\frac{{B_{z}}^{2}}{2}$. Physically this is in accordance with the
fact that the radial stresses , here including the magnetic stress
$(T^{1}_{1})$ must vanish at the cylinder surface. The heat flow
also vanishes which in the ECM field equations implies that the RC
rotation must be in principle constant. Note that the signs of the
pressures along orthogonal directions indicate that the magnetic
string undergoes gravitational collapse which in turn helps dynamo
mechanism. Physical applications of the model discuss here may be in
the investigation of the gravitational extra effects on the
well-known Einstein-de Haas effect due to the non-Riemannian effects
from the spin density. To use this Einstein-de Haas idea we may
simply cancel the term in front of ${d{\phi}}^{2}$. This yields
\begin{equation}
B^{2}={{\sigma}_{0}}^{2}R^{4} \label{43}
\end{equation}
whose square root yields
\begin{equation}
B=\pm{{\sigma}_{0}}R^{2} \label{44}
\end{equation}
Note that when the string radius is the radius of an H atom,
$R_{H}\sim{10^{-8}cm}$ this yields
\begin{equation}
B=\frac{c^{3}T}{8{\pi}G}R^{2} \label{45}
\end{equation}
when $G=G_{f}\sim{10^{30}cgs units}$ is the f-meson gravity
dominance gravitational constant, this yields
\begin{equation}
B=10^{-16}{\times}R^{2} \label{46}
\end{equation}
which in turn yields
\begin{equation}
B=10^{-32} Gauss \label{47}
\end{equation}
In the case of superconducting cosmic strings \cite{8,9} the
relation between electric current J and magnetic field is given by
the Biot-Savart law for the magnetic cylinder
\begin{equation}
B=\frac{2J}{R} \label{48}
\end{equation}
which immeadiatly tells us that the electric current over the string
is extremely small. As shown by Witten \cite{9} in general
superconducting cosmic strings possesses strong electric currents
and not weak like that. Note that earlier D Battefeld et al
\cite{10} have shown that by investigating the magnetogenesis and
show that $B\sim{10^{-29}G}$ has been obtained on
$5\rightarrow{5kpc}$ are needed to account for magnetic fields in
spirals galaxies, accounts using cosmic strings; during the collapse
of protogalactic clouds. This value is only 3 orders of magnitude we
obtained here. Actually gravitational wavces and dragging effects
leads to $10^{-28}Gauss$.
\section{Spin-torsion polarised cosmic strings and large scale
magnetic fields} In this section we shall consider a rotating string
instead of non-rotating string of the last section. Let us now
consider the spin-polarised cylinder metric
\begin{equation}
ds^{2}=-[exp(\alpha)dt+Md{\phi}]^{2}+r^{2}exp(-\alpha)d{\phi}^{2}+exp(2\beta-2\alpha)[dr^{2}+dz^{2}]
\label{49}
\end{equation}
as given in Soleng \cite{4}. By making use of ECM equations this
reduces to
\begin{equation}
\frac{k{B_{z}}^{2}}{2}={\Omega}^{2} \label{50}
\end{equation}
where ${\Omega}^{2}$ is the square of vorticity of the
spin-polarised cylinder. The remaining ECM equations are
\begin{equation}
k\rho-\frac{k{B_{z}}^{2}}{2}={\Omega}^{2}-2{\sigma}{\Omega}
\label{51}
\end{equation}
\begin{equation}
kp_{z}+\frac{k{B_{z}}^{2}}{2}=-{\Omega}^{2}+2{\sigma}{\Omega}
\label{52}
\end{equation}
Here ${\sigma}$ is the spin-density. Let us now express the
cylindrical string as
\begin{equation} ds^{2}=
-[dt+k(k{\sigma}^{2}-\frac{{B_{z}}^{2}}{\sqrt{2}})r^{2}d{\phi}^{2}]^{2}+r^{2}d{\phi}^{2}+dr^{2}+dz^{2}
\label{53}
\end{equation}
Thus magnetic energy to ${B_{z}}^{2}$ is proportional to spin
density ${\sigma}^{2}$. Thus one obtains
\begin{equation}
B_{z}\sim{\sigma}\label{54}
\end{equation}
This formula is obtained by the cancelment of the term between
spin-torsion effects and magnetic fields as in Einstein-de Haas
effect. Since the torsion scalar T is given by
\begin{equation}
T=\frac{4{\pi}G}{c^{3}}{\sigma} \label{55}
\end{equation}
Now let us show the physical importance of this correction by
showing that inversion of this last expression which yields
\begin{equation}
{\sigma}=\frac{c^{3}}{4{\pi}G}T \label{56}
\end{equation}
and substitution in the now corrected expression (\ref{2}) yields an
important expression between the magnetic field and torsion already
discussed in previously to estimate magnetic fields obtained from
torsion fields \cite{56}. Thus
\begin{equation}
B_{z}\sim{\frac{c^{3}}{4{\pi}G}T} \label{57}
\end{equation}
By making use of the torsion field on earth laboratory as given in
Hughes-Drever experiment by Laemmerzahl \cite{5} as
$T\sim{10^{-17}cm^{-1}}$, and the strong gravity gravitational
constant \cite{2} $G_{f}=10^{30}cgs units$ one obtains the following
magnetic field $B\sim{10^{-22}Gauss}$ which as we wish to prove very
well within the magnetic field to seed galactic dynamos as shown
recently by Barrow et al \cite{13}. Let us now consider the
cosmological magnetic field obtained from the ECM equations in terms
of the rotation the first ECM equations is
\begin{equation}
B_{z}=2{\frac{1}{\sqrt{k}}}{\Omega} \label{58}
\end{equation}
where ${\Omega}$ is the Harrison-Rees vorticity which for galaxies
is $10^{-9}rad s^{-1}$. From the above value for Einstein´s
gravitational constant one obtains $B_{z}\sim{10^{-5}Gauss}$ which
is the intergalactic magnetic field. Note that investigation of
magnetogenesis \cite{11} in Einstein-Cartan gravity is not a novel
subject but in the previous works \cite{12} much more attention has
been given to torsion than to metric effects like in this paper.
Note that cosmic strings we have addressed here are
non-superconducting contrary to ones investigated by Witten and more
recently by Dimopoulus and Davies \cite{14} with friction and
backreaction.

\begin{flushleft}
{\large Acknowledgements}
\end{flushleft}
I would like to thank Professors H.Soleng, T Battefeld and
P.S.Letelier for helpful discussions on the subject of this paper.
Financial support from CNPq. and UERJ is gratefully acknowledged.

\end{document}